\documentclass[usenatbib]{mnras}

\usepackage{newtxtext,newtxmath}

\usepackage[T1]{fontenc}
\usepackage{ae,aecompl}

\usepackage{graphicx}	
\usepackage{amsmath}	
\usepackage{amssymb}	

\newcommand{\hmpc}{\,$h^{-1}$\,Mpc}
\newcommand{\hmsun}{\,$h^{-1}\,M_{\odot}$}
\newcommand{\reff}{$R_{\rm eff}$}
\newcommand{\lcdm}{$\Lambda$CDM}

\title[Vainshtein Voids]{Using Voids to Unscreen Modified Gravity}
\author[Falck, et al.]{Bridget Falck,$^1$\thanks{E-mail:bridget.falck@astro.uio.no}
	Kazuya Koyama,$^2$ 
	Gong-Bo Zhao,$^{3,2}$ 
	Marius Cautun$^4$ \\
	$^1$Institute of Theoretical Astrophysics, University of Oslo, PO Box 1029 Blindern, N-0315, Oslo, Norway \\
	$^2$Institute of Cosmology and Gravitation, University of Portsmouth, Dennis Sciama Building, Burnaby Rd, Portsmouth PO1 3FX, UK \\
    $^3$National Astronomy Observatories, Chinese Academy of Science, Datun Road, Beijing, PR China \\
	$^4$Institute for Computational Cosmology, Department of Physics, Durham University, Durham DH1 3LE, UK}
\date{Draft version \today}

\pubyear{2017}

\begin{document}
\label{firstpage}
\pagerange{\pageref{firstpage}--\pageref{lastpage}}
\maketitle

\begin{abstract}
The Vainshtein mechanism, present in many models of gravity, is very effective at screening dark matter halos such that the fifth force is negligible and general relativity is recovered within their Vainshtein radii. Vainshtein screening is independent of halo mass and environment, in contrast to e.g. chameleon screening, making it difficult to test. However, our previous studies have found that the dark matter particles in filaments, walls, and voids are not screened by the Vainshtein mechanism. We therefore investigate whether cosmic voids, identified as local density minima using a watershed technique, can be used to test models of gravity that exhibit Vainshtein screening. We measure density, velocity, and screening profiles of stacked voids in cosmological $N$-body simulations using both dark matter particles and dark matter halos as tracers of the density field. We find that the voids are completely unscreened, and the tangential velocity and velocity dispersion profiles of stacked voids show a clear deviation from \lcdm\ at all radii. Voids have the potential to provide a powerful test of gravity on cosmological scales.

\end{abstract}



\section{Introduction}

The concordance cosmological model, \lcdm, in which a cosmological constant drives the late-time accelerated expansion of the Universe, is currently in agreement with cosmological observations. However, the predicted value of the vacuum energy is many orders of magnitude larger than the observed value of the cosmological constant -- the so-called cosmological constant problem~\citep[see, e.g.,][]{Carroll1992}. Theoretical approaches to modeling the late-time acceleration generally follow one of two avenues: either the acceleration is caused by a dark energy field, which need not have a constant energy density, or a new theory of gravity is needed which modifies general relativity (GR) on large scales. Theories which modify GR are distinguished by their screening mechanism, which describes the transition from small scales where GR is well-tested to large scales where modifications result in accelerated expansion~(for a detailed review, see \citep{Joyce2015, Koyama2016}).

Modifications to GR generally introduce a new scalar degree of freedom which mediates a fifth force, and screening mechanisms suppress this fifth force on small scales. For example, the chameleon mechanism makes the mass of the scalar field large in high density environments~\citep{Khoury2004}, while the symmetron and dilaton mechanisms change the scalar field coupling to matter~\citep{Hinterbichler2010,Brax2010}. Screening also occurs if the derivative self-interactions of the scalar field become large, which is realized for k-mouflage~\citep{Babichev2009,Brax2014}, D-BIonic~\citep{Burrage2014}, and Vainshtein~\citep{Vainshtein1972} screening mechanisms. The Vainshtein mechanism is particularly interesting because it appears in a large class of modified gravity theories such as massive gravity~\citep{deRham2014,Koyama2011,Sbisa2012}, galileon cosmology~\citep{Chow2009,Silva2009}, and the DGP braneworld model~\citep{DGP2000,Maartens2010}.

In this paper, we investigate the effect of the Vainshtein mechanism on cosmic voids. Cosmic voids are hierarchical underdense regions of the universe marked by outflow from void centers to nearby structures, slowly expanding as halos and filaments collapse~\citep[see, e.g.][]{Aragon2013,Nadathur2015Voids1,Sutter2014}. Their general features can be understood theoretically via the excursion set formalism, in a similar way to halos~\citep{Sheth2004,Jennings2013}. These theoretical models aren't perfect: it turns out that void boundaries do not usually correspond to regions of shell-crossing~\citep{Nadathur2015Voids1,Falck2015Voids,Achitouv2015}, and the models do not account for the non-spherical nature of voids found by watershed methods of e.g.~\citet{Platen2007,Neyrinck2008}. Nevertheless, the non-sphericity of voids is washed out when many voids are averaged together; the Alcock-Paczynski test can thus be used to measure cosmological parameters from stacked voids~\citep{Lavaux2010,Sutter2012}, and the stacked density profiles seem to be self-similar~\citep{Hamaus2014,Ricciardelli2014,Nadathur2015,Cautun2016}. 

Screening mechanisms typically involve nonlinear dynamics in the equation of motion for the scalar field. Thus cosmological $N$-body simulations that solve for the nonlinear gravitational collapse of structures are required to compare these models to GR and search for observable signatures; see~\citet{Winther2015Code} for a review and comparison of such codes. 
Since screening mechanisms in modified gravity models operate in high density regions, dark matter halos are often screened, so it can be hard to detect deviations from GR using galaxies or clusters. 
Indeed, previous simulations of the Vainshtein mechanism models have found that it is more efficient at screening dark matter halos than other types of screening~\citep{Schmidt2010,Li2013Galileon,Barreira2013,Barreira2014,Falck2014,Falck2015Screening}, though there may be signatures of these models in the velocity field~\citep{Lam2012,Hellwing2014, Bose2017} and higher order hierarchical amplitudes \citep{Hellwing2017}. 

This makes voids a potentially fruitful tool for probing the nature of gravity and the accelerated expansion. Indeed, it has recently been proposed that redshift-space distortions around voids can provide precise measurements of the growth rate of structure, thereby probing deviations from GR~\citep{Hamaus2016,Cai2016}. The excursion set formalism has been extended to predict the abundance of voids in chameleon and symmetron models of gravity~\citep{Clampitt2013,Lam2015,Voivodic2017}, and their properties have been studied in simulations of chameleon, symmetron, and Galileon models~\citep{Li2012,Cai2015,Zivick2015,Barreira2015,Hamaus2015,Voivodic2017}. Similarly to voids, troughs are underdense regions along the line-of-sight in galaxy surveys~\citep{Gruen2015}, and their weak lensing signal has been studied in simulations of the normal branch of DGP gravity~\citep{Barreira2017Troughs}. 
This paper presents the first study of voids in the Vainshtein mechanism using simulations of the normal-branch DGP model. This has the same expansion history as the \lcdm\ model so that we can disentangle the effects of the background model from those of the Vainshtein mechanism.  Voids have also been studied in the Cubic Galileon model, which exhibits the Vainshtein screening mechanism, but this model suffers from an instability in underdense regions at late times such that the quasi-static solution ceases to exist~\citep{Barreira2015,Winther2015}; the nDGP model is free of this problem.

This paper proceeds as follows. We present the basic theory in Section~\ref{sec:theory} and describe the simulations and void identification method in Section~\ref{sec:methods}. Results are given in Section~\ref{sec:results}; we compare the distributions of voids in GR to those in the Vainshtein mechanism, as well as the density, fifth force, and velocity profiles, at $z=1$ and $z=0$. Conclusions are given in Section~\ref{sec:conc}.

\section{Theory}
\label{sec:theory}

\subsection{Model}
We consider the normal branch DGP (nDGP) braneworld model that has exactly the same expansion history as the \lcdm\ model. Under the quasi-static approximation, the Poisson equation and the equation for the scalar field $\varphi$ are given by~\citep{Koyama2007}
\begin{align}
\nabla^2 \Psi & = \nabla^2 \Psi_N + \frac{1}{2} \nabla^2 \varphi, \\
\nabla^2 \varphi & + \frac{r_c^2}{3 \beta(a) a^2}
[(\nabla^2 \varphi)^2 -(\nabla_i \nabla_j \varphi)(\nabla^i \nabla^j \varphi)]
= \frac{8 \pi G a^2}{3 \beta(a)} \rho \delta,
\label{eq:phievo1}
\end{align}
where $\Psi$ is the gravitational potential and $\nabla^2\Psi_N= 4 \pi G a^2 \rho \delta$. The scalar field $\varphi$ mediates an additional ``fifth force''. 
Note that the quasi-static approximation has been shown to have a negligible effect on results~\citep{Winther2015}. The function $\beta(a)$ is given by
\begin{equation}
\beta(a) = 1 +  2 H r_c \left(1+ \frac{\dot{H}}{3 H^2} \right),
\label{eq:beta}
\end{equation}
where $r_c$ is the cross-over scale, which is a free parameter of the model. Note that $\beta$ is always positive, so the growth of structure formation is enhanced in this model.

As mentioned, this model has one extra parameter, $r_c$, in addition to the usual cosmological parameters in the \lcdm\ model. If $r_c$ increases, the enhancement of gravity becomes weaker, the Vainshtein mechanism operates more efficiently, and we recover \lcdm.

\subsection{Voids}

In order to obtain analytic predictions for the forces in voids, we assume the density profile can be described by~\citep{Hamaus2014}
\begin{equation}
\delta(R'=R/R_{\rm eff}) = \delta_v \frac{1 - (R'/s_1)^{\alpha}}{1 + 
	(R'/s_2)^{\beta}}. 
\label{eq:profile}
\end{equation}
This admits an analytical formula for the mass perturbation 
$M(<R) = 4 \pi \bar{\rho}_m \int^R_0 \delta(x) x^2 dx$ in terms of the hypergeometric functions. 

The Newtonian force is given by
\begin{equation}
\frac{d \Psi_N}{d R} = \frac{G M(<R)}{R^2}. 
\end{equation}
The scalar field equation in the nDGP model, Eq.~(\ref{eq:phievo1}) can be solved analytically  
\begin{equation}
\frac{d\varphi}{dR} = \frac{G M(<R)}{R^2} \frac{4}{3 \beta} g\left(\frac{R}{r_*} \right), \quad
g(x) = x^3 \left( \sqrt{1+x^{-3}} -1 \right),
\end{equation}
where $r_*$ is the Vainshtein radius 
\begin{equation}
r_*^3= \frac{16 G M(<r) r_c^2}{9 \beta^2}.
\label{r*}
\end{equation}
Note that $r_*^3$ is negative for voids thus $x = R/r_*$ is negative. If $x$ becomes smaller than $-1$, the inside of the square root in $g(x)$ becomes negative and the solution ceases to exist. This happens in galileon models~\citep{Barreira2015}, and it was shown that this problem does not go away even we include the time dependence in the scalar field~\citep{Winther2015}. However, this problem does not occur in the nDGP model. The condition that $x > -1$ is satisfied for an empty void with $\delta =-1$ is given by 
\begin{equation}
\frac{9}{8} \frac{\beta a^3}{\Omega_m (H_0 r_c)^2} >1 
\end{equation} 
where $\Omega_m$ is the density parameter for matter and $H_0$ is the present day Hubble parameter. This condition is always satisfied with $\beta$ given by Eq.~(\ref{eq:beta})~\citep{Winther2015}. 


\section{Methods}
\label{sec:methods}

\subsection{Simulations}
\label{sec:sims}

We run cosmological $N$-body simulations of the nDGP model and \lcdm\ using the AMR code ECOSMOG~\citep{Li2012code,Li2013}, which is a modified gravity version of RAMSES~\citep{ramses}. The background cosmology is taken from WMAP9~\citep{wmap9}: $\Omega_m = 0.281$, $h=0.697$, and $n_s=0.971$. The simulations have a box of length of 1024\hmpc, $1024^3$ dark matter particles, a starting redshift of $49$, and the initial conditions were generated using {\tt MPGrafic}~\citep{mpgrafic}
\footnote{Available at \url{http://www2.iap.fr/users/pichon/mpgrafic.html}}.

We run two nDGP simulations with different values of the cross-over scale and different values of $\sigma_8$. These values were chosen to match $f(R)$ simulations with the same $\sigma_8$ at $z=0$, such that nDGP2 is matched to F5 ($|f_{R0}|=10^{-5}$) and nDGP3 is matched to F6 ($|f_{R0}|=10^{-6}$), thus nDGP2 deviates more strongly from \lcdm\ than nDGP3. Specifically, for nDGP2, $H_0 r_c = 0.75$ and $\sigma_8 = 0.902$, and for nDGP3, $H_0 r_c = 4.5$ and $\sigma_8 = 0.859$. The \lcdm\ simulation has $\sigma_8 = 0.844$. Note that these values differ from those in our previous papers~\citep{Falck2014,Falck2015Screening} because of the difference in the background cosmology; note also that we do not simulate nDPG1 (corresponding to F4) because of the intense computational requirements and the observational constraints already present~\citep{Joyce2015}.

\begin{figure}
\includegraphics[width=\hsize]{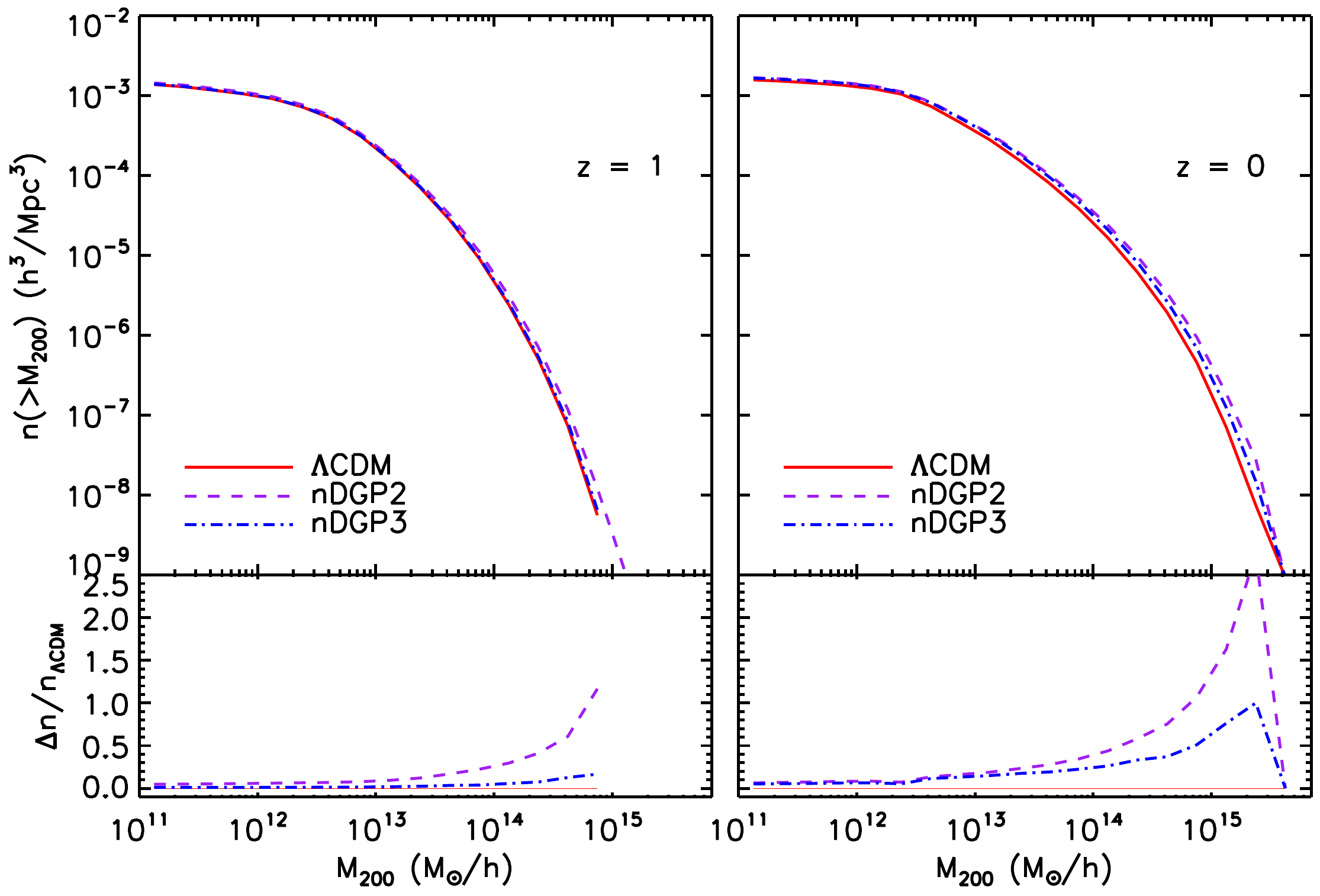}
\caption{$M_{200}$ mass functions of ROCKSTAR halos at $z=1$ (left panel) and $z=0$ (right panel) and the ratio of nDGP to \lcdm\ mass functions (bottom panels).}
\label{fig:massfn}
\end{figure}

We identify halos using ROCKSTAR~\citep{Behroozi2013}, a phase-space halo finder. The $z=1$ and $z=0$ mass functions are shown in Figure~\ref{fig:massfn} for \lcdm\ and the two nDGP models, and ratios of nDGP to \lcdm\ mass functions are shown in the bottom panels. At $z=1$, there are many more halos with $M_{200}>10^{14}$\hmsun\ in nDGP2 than \lcdm, and slightly more large halos in nDGP3. This difference increases at $z=0$; both nDGP2 and nDGP3 have $\sim 20\%$ more halos with $M_{200}>10^{13}$\hmsun, and nDGP2 has over 100\% more halos with masses greater than $10^{15}$\hmsun, though the mass functions at the highest mass bins are likely affected by small number statistics. This behavior is a reflection of the fact that the growth of structure formation is stronger in these models.


\subsection{Void Identification}
\label{sec:voidmethod}

We use a watershed technique to identify voids in the simulations~\citep{Platen2007}, in which voids are local density minima and void boundaries are the higher density ridges between them. The watershed technique can be used to identify voids given any set of discrete tracers of the underlying density field and has been successfully applied to define void catalogs in galaxy surveys~\citep{Sutter2012DR7,Nadathur2014DR7,Nadathur2016DR11,Mao2016}. We measure the density field using the Delaunay Tessellation Field Estimator~\citep{Schaap2000,vdW2009,Cautun2011} which constructs a volume-weighted density at the locations of the discrete tracers, which we take to be either the dark matter particles or the halos identified with ROCKSTAR. 
The mean separation of the density tracers determines the scale at which local density minima can be resolved and thus the size of the smallest voids; note that due to its adaptive nature, DTFE is less susceptible to shot noise at low densities than, e.g., cloud-in-cell or other grid-based measures of the density field. For these simulations, the mean density of dark matter particles is 1 per cubic \hmpc, and for the halos it is roughly $1.5\times 10^{-3}$ per cubic \hmpc. 

This DTFE density field, defined either by dark matter particles or by halos, is interpolated onto a grid of cell size 1\hmpc\ for computational convenience and then smoothed with a Gaussian filter of size 2\hmpc\ to reduce spurious voids caused by shot noise. The watershed algorithm then defines the void boundaries that separate local density minima. In principle, the watershed method can identify a hierarchy of sub-voids within larger voids, but we do not consider sub-voids here. We require that each cell volume is part of only one void, and boundaries must have a density contrast of at least $\delta = -0.8$. 

Profiles are calculated by averaging quantities in spherical shells around the barycenter, which is the volume-weighted average position of the grid cells that make up the void. For voids found using dark matter particles as tracers of the density field, we average quantities (density, velocity, or force) of the particles, and for voids identified in the halo distribution we measure profiles using the positions and velocities of the halos. Thus halo-identified voids will have profiles that are not as well resolved but are closer to what can be measured in a galaxy survey. Bringing simulation results even closer to observations requires the use of mocks to take into account the bias of galaxy tracers of the density field~\citep{Ricciardelli2014,Nadathur2015Voids2}.


\section{Results}
\label{sec:results}

\subsection{Void Volume Functions}

\begin{figure}
\includegraphics[width=\hsize]{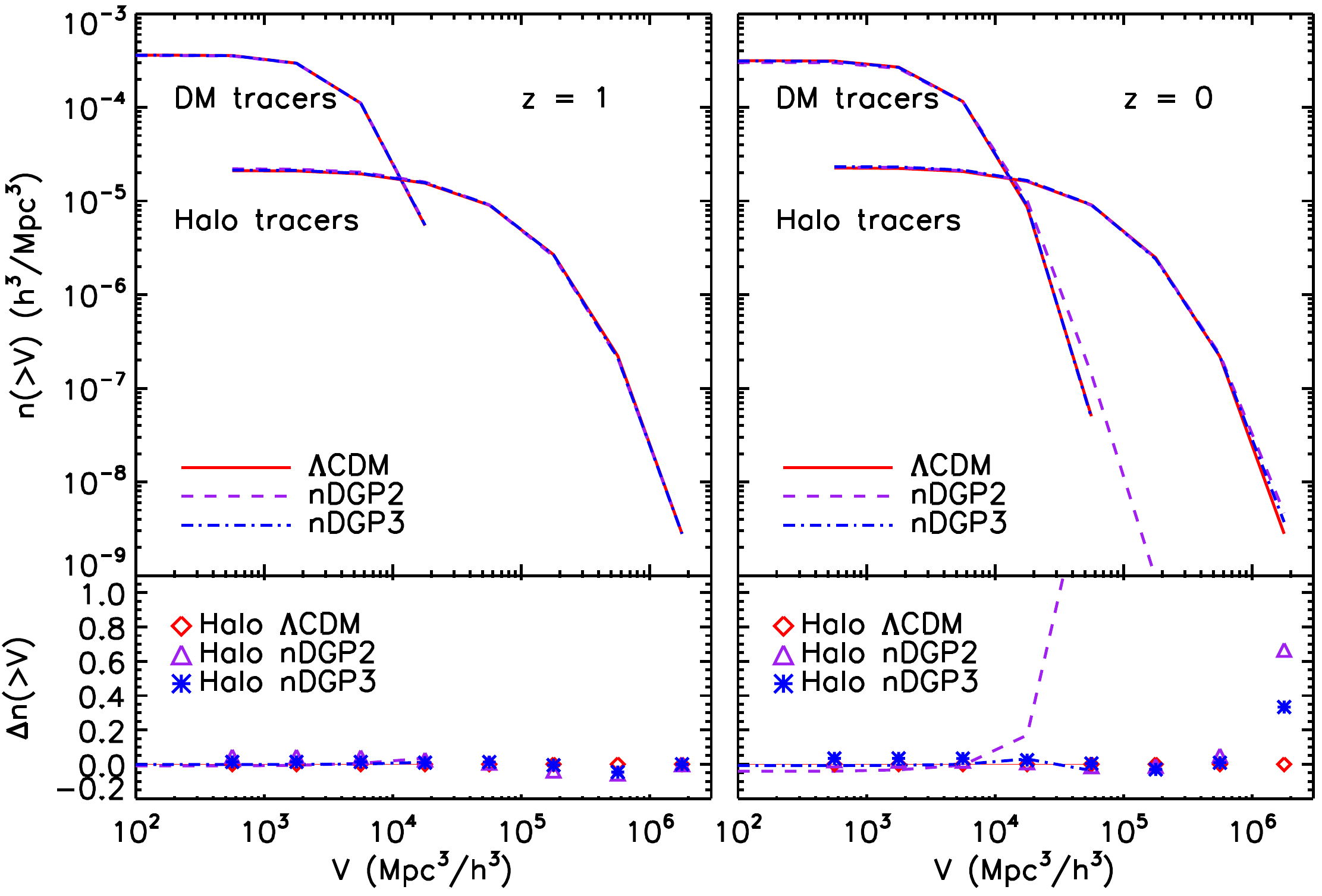}
\caption{Void volume functions at $z=1$ (left panel) and $z=0$ (right panel) and the ratio of nDGP to \lcdm\ volume functions (bottom panels). In the bottom panels, lines indicate voids found using dark matter particles, and symbols indicate voids found in the distribution of halos.}
\label{fig:volfn}
\end{figure}

In Figure~\ref{fig:volfn} we show the cumulative distribution functions of void volumes at $z=1$ and $z=0$, for voids found in all models using both dark matter particles and halos as tracers of the density field. Since halos are sparser than dark matter particles and are biased tracers of the density field, there are fewer voids found using halos as tracers, and these voids are much larger~\citep{Ricciardelli2014,Nadathur2015Voids1}. The distributions change very little from $z=1$ to $z=0$, but there are more large voids at $z=0$ than $z=1$, as voids slowly grow and their interiors evacuate~\citep{Sutter2014}.

It is clear from the figures that there is little difference between the distribution of voids in the \lcdm\ and nDGP simulations; in most of the volume bins the nDGP cumulative volume functions remain very close to that of the \lcdm\ simulation, as seen in the ratios in the bottom panels of Figure~\ref{fig:volfn}. At $z=0$, there are more large voids in the nDGP2 simulation than \lcdm\ and nDGP3. A plausible reason for this is that since gravity is stronger in this model, voids can evacuate more quickly, which can prevent large voids from being identified as several smaller voids. Note that at $z=0$, there are $\sim 50$ voids (traced by dark matter particles) in the largest volume bin of \lcdm\ and nDGP3 and $\sim 150$ nDGP2 voids (purple dashed line) in the same bin, suggesting this excess of large voids may be significant. However, there are fewer than 5 \lcdm, nDGP2, and nDGP3 halo-tracer voids in the largest volume bin; larger simulations are needed to study in detail the differences in the populations of very large voids. Further, we will see that this excess of large voids has little bearing on the profiles of stacked voids measured in the next sections.

Void volumes are often described in terms of their effective radius, \reff, defined as the radius of a sphere having the same volume as the void, \reff$=(3V/4\pi)^{1/3}$, even though the voids themselves can be non-spherical.  For all models and at both redshifts, the distributions of void sizes peak at $\sim 8$\hmpc\ for voids found using dark matter particles as tracers and at $\sim 15$\hmpc\ for voids found using halo tracers. In what follows, we split the voids into two samples according to these median values in order to take into account the physical and dynamical differences between small and large voids. In our simulations, there are $\sim 20000$ small and $\sim 19000$ large voids using dark matter particle tracers, and there are $\sim 8000$ small and $\sim 15000$ large voids using halo tracers.

\subsection{Density profiles}

\begin{figure*}
\parbox{0.45\textwidth}{\includegraphics[width=\hsize]{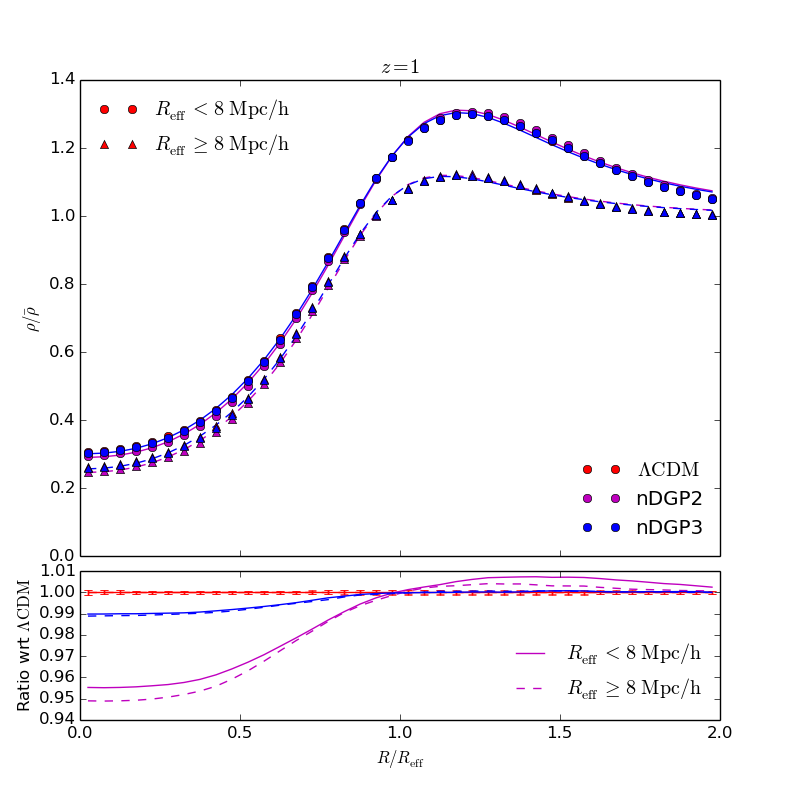}}%
\qquad
\begin{minipage}{0.45\textwidth}%
\includegraphics[width=\hsize]{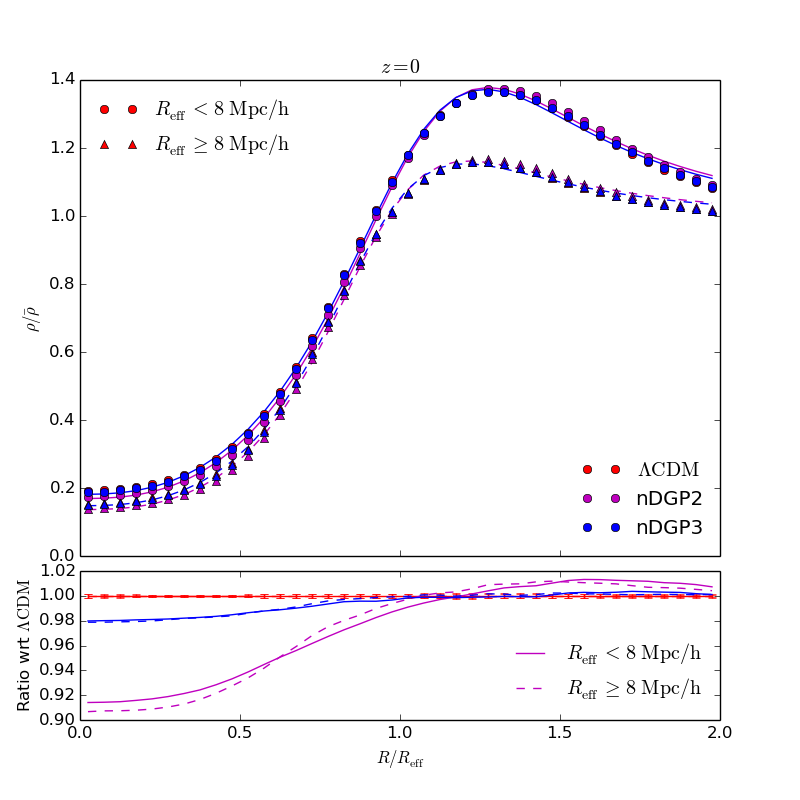}
\end{minipage}%
\caption{Mean density profiles of voids found in the dark matter particle distribution at $z=1$ (left panel) and $z=0$ (right panel). In the upper panels, symbols give results from simulations, and the lines are the best-fit analytic profiles using the fitting function of Equation~\ref{eq:profile}. In the bottom panel, ratios of nDGP to \lcdm\ profiles from the simulations are given as solid and dashed lines, and (very small) error bars denoting the error on the mean are shown for \lcdm\ only.}
\label{fig:denprofs}
\end{figure*}

We plot the density profiles of stacked voids found using dark matter particles as tracers at $z=1$ and $z=0$ in Figure~\ref{fig:denprofs}. There is a clear difference between the profiles of large and small voids; the small voids are shallower and have a more compensated profile, since these voids tend to live in dense environments, while the larger voids are deeper and their profiles have a less prominent density ridge at the boundary~\citep{Sheth2004,Hamaus2014,Cautun2016}. The difference between void density profiles in \lcdm\ and the two nDGP simulations are not as clear as the difference between the profiles of small and large voids, so we also plot the ratios of the nDGP to \lcdm\ density profiles in the bottom panels of Figure~\ref{fig:denprofs}, where error bars represent the standard deviation of all void profiles contributing to the stack and are shown for \lcdm\ only. The model with the strongest deviation from GR, nDGP2, has a deeper stacked profile and a correspondingly slightly more pronounced ridge, but these differences are less than 5\% at $z=1$. From $z=1$ (left panel) to $z=0$ (right panel), the density profiles become emptier in the centers and larger at the void edges, as matter continues to evacuate from void centers and pile up at void boundaries, and the difference between the nDGP2 and \lcdm\ profile centers increases to 10\%.

The stacked density profiles are fit to a 5-parameter analytic model (see Equation~\ref{eq:profile}), and these fits are shown as solid and dashed lines for the small and large void samples, respectively, in the upper panels only. These analytical fits are used to calculate the force profiles in the following sections.

\begin{figure}
\includegraphics[width=\hsize]{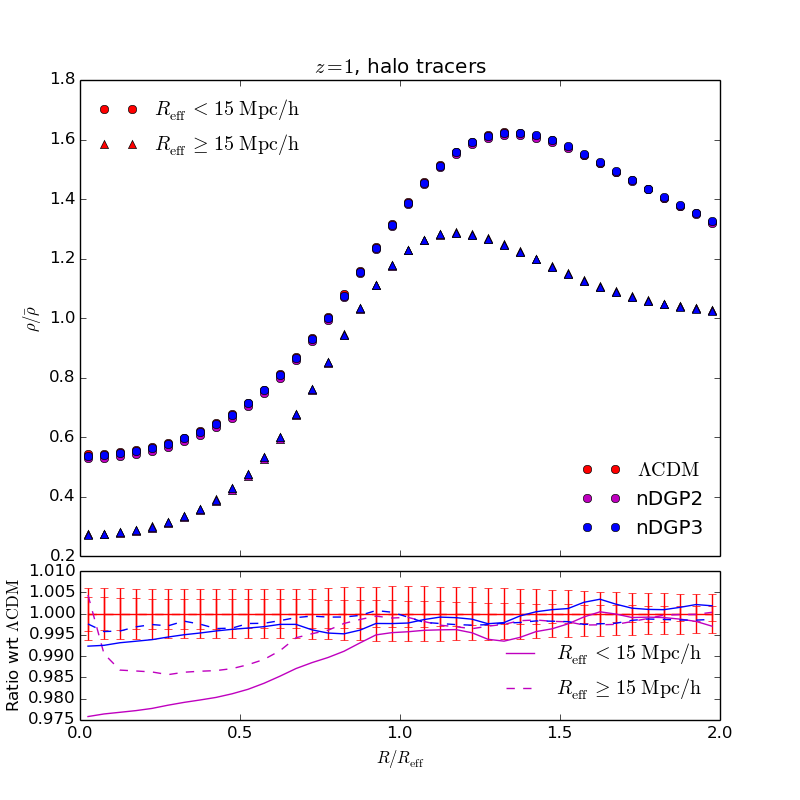}
\caption{Density profiles: the same as Figure~\ref{fig:denprofs} but for voids found using halo tracers at $z=1$. The error bars, again shown for \lcdm\ only, are much larger because there are fewer voids found in the halo distribution, and for the same reason the error bars are larger for the set of small voids (both are shown).}
\label{fig:haloden08}
\end{figure}

When halos are used as tracers of the density field, both to define the voids and to measure the profiles, the stacked density profiles again show much more dependence on void radius than on the gravity model, as shown in Figure~\ref{fig:haloden08} for $z=1$. Compared to the voids found in the dark matter, these voids are more dense both in the void centers and the void edges, especially the sample of ``small'' voids (which are larger than those found using dark matter particles). There is more of a difference between the small and large void density profiles, likely because halos are biased tracers of the density field; we note that measuring profiles in galaxy surveys would result in somewhat different profiles due to galaxy bias~\citep{Nadathur2015Voids2}. The ratio of nDGP to \lcdm\ profiles are again shown in the bottom panel of Figure~\ref{fig:haloden08}; since there are fewer voids, the error bars are much larger than for voids found using dark matter particles, but the general trend appears to be the same, with the nDGP stacked profiles being emptier than in \lcdm. Emptier voids are also found in studies of $f(R)$ and Galileon models~\citep{Cai2015,Barreira2015}, thus they are a common feature of models in which a fifth force enhances gravity, but Figure~\ref{fig:haloden08} shows that it will be difficult to use density profiles of voids to test Vainshtein screening.

\subsection{Screening profiles}

In dark matter halos, the fifth force is suppressed within the Vainshtein radius by the Vainshtein screening mechanism. The Vainshtein mechanism has been shown to be very efficient at screening halos regardless of their mass, the density of their local environment, or their location within the cosmic web~\citep{Schmidt2010,Falck2014,Falck2015Screening}. However, due to the dimensional dependence of the nonlinear equations describing Vainshtein screening~\citep{Bloomfield2015}, the Vainshtein mechanism exhibits a shape dependence and does not work for objects that are not collapsing along three dimensions such as voids, walls, and filaments~\citep{Falck2014}. When the Vainshtein mechanism is not working, the ratio of the fifth force to the Newtonian force, 
\begin{equation}
\Delta_M = \frac{1}{2} \frac{d \varphi/ dr} {d \Psi_N/dr},
\end{equation}
has the linear theory value of $\Delta_M = 1/3 \beta$ -- this is where we expect to find the largest signals for theories which contain the Vainshtein screening mechanism. In this section, we measure the radial profiles of the fifth force and Newtonian force of voids found in the dark matter distribution, as well as their ratio, using the forces saved during the simulation run for each dark matter particle. Note that this means if particles are moving away from void centers, they will have positive radial forces, as expected for underdense, expanding voids.

\begin{figure*}
\parbox{0.45\textwidth}{\includegraphics[width=\hsize]{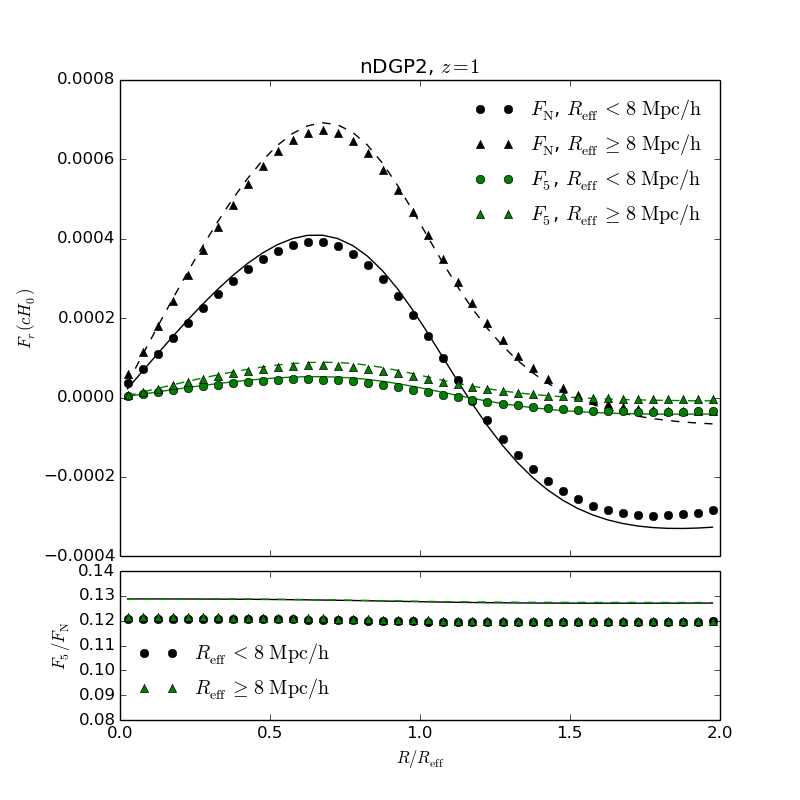}}%
\qquad
\begin{minipage}{0.45\textwidth}%
\includegraphics[width=\hsize]{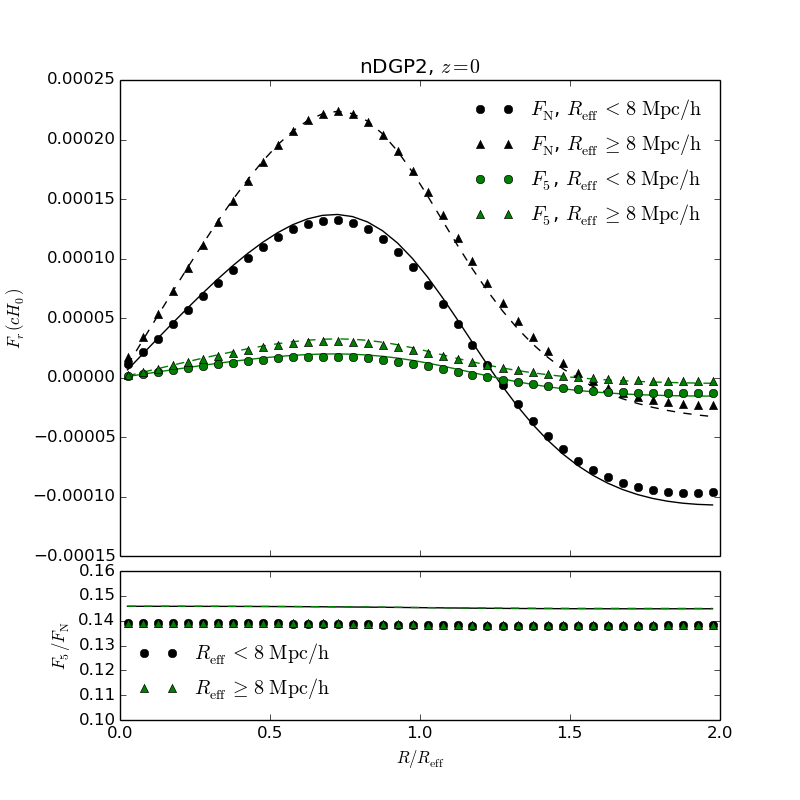}
\end{minipage}%
\caption{Radial profiles of the fifth force (green) and Newtonian force (black) of voids found using dark matter particles for nDGP2 at $z=1$ (left panel) and $z=0$ (right panel). Forces are in units of $c H_0$. Symbols give results from the simulation, while solid and dashed lines are analytic predictions using the density profile fits for small and large voids, respectively. Bottom panels show profiles of the ratio of the fifth to Newtonian force, $\Delta_M$.}
\label{fig:fraddgp2}
\end{figure*}

Figure~\ref{fig:fraddgp2} shows the $z=1$ and $z=0$ stacked force profiles for the voids in the nDGP2 simulation. The larger voids (triangles) have profiles that increase from the void center and gradually return to zero at large radii, while the smaller voids (circles) have radial force profiles that go to zero at the effective radius and negative at larger radii as the voids are squeezed by the collapse of their local neighborhood, creating the compensated density profiles shown in the previous section. The fifth force profiles show a similar behavior to the Newtonian force, though with a much smaller magnitude. Analytic predictions for the fifth and Newtonian forces are calculated using the fits to the density profiles and shown as solid and dashed lines for small and large voids, respectively.  

The force ratios, $\Delta_M$, are shown in the bottom panels of Figure~\ref{fig:fraddgp2}. They are constant with radius and have values very near the linear theory prediction of 0.127 at $z=1$ and 0.145 at $z=0$, for both the large and small void samples. Note that the analytic $\Delta_M$ profiles overpredict the value measured from simulations by a constant amount because they are calculated from the stacked density profiles: averaging the matter field suppresses higher density peaks and a smoother and lower density void profile is obtained, as discussed in~\citet{Barreira2015}. The fifth force derived from this lower density profile is thus higher, since the screening is underestimated. The analytical calculation that uses this profile gives a fifth force that is stronger in magnitude than that obtained by averaging the force field directly.

Figure~\ref{fig:fraddgp2} shows that, unlike halos, cosmic voids are completely unscreened in the Vainshtein mechanism. Even going out to twice the void effective radii, the stacked profiles of voids remain unscreened. This is likely because most of the density ridges that make up void boundaries can be classified dynamically as filament, wall, and void components of the cosmic web~\citep{Falck2015Voids}, which are unscreened~\citep{Falck2014}, instead of halos, which are screened but occupy a very small volume.

For voids found using halos as tracers, however, more of the density ridges that make up void boundaries can be expected to contain screened halos. One might expect this to introduce a radial dependence to the screening profiles, such that screening is suppressed near void boundaries, but we will see in the next section that no such radial dependence is found for velocity dispersion profiles of halo-tracer voids, which trace the fifth force. This is because, though dark matter halos themselves are screened in the Vainshtein mechanism, screened objects can still feel the fifth force of external fields~\citep{Hui2009,Falck2014}, so screened halos that trace voids can still be influenced by the dynamics of external fields if their wavelengths are long compared to the Vainshtein radius of the halos.

\begin{figure*}
\parbox{0.45\textwidth}{\includegraphics[width=\hsize]{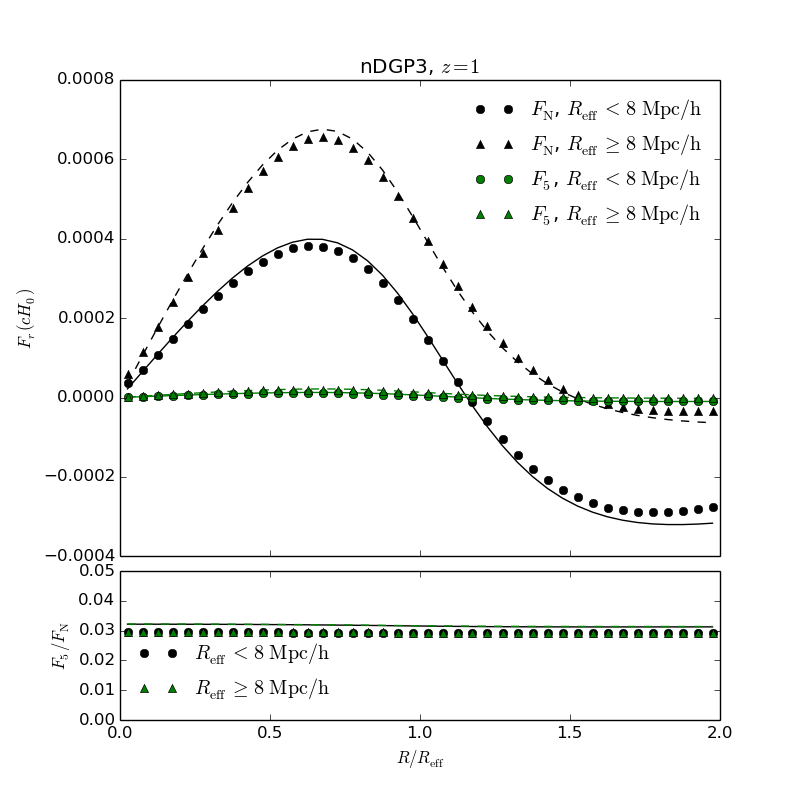}}%
\qquad
\begin{minipage}{0.45\textwidth}%
\includegraphics[width=\hsize]{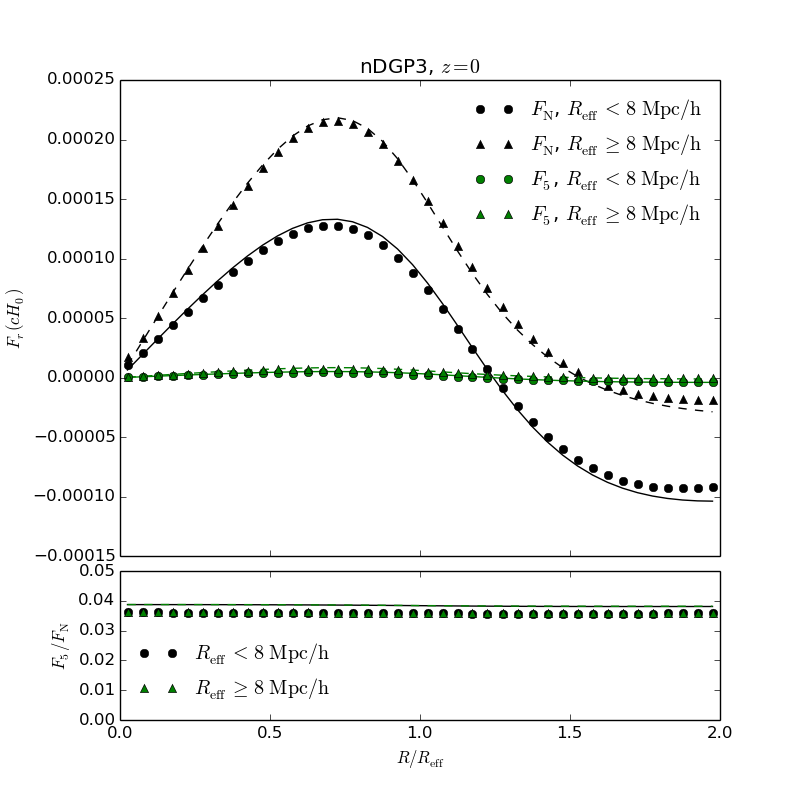}
\end{minipage}%
\caption{Force profiles: the same as Figure~\ref{fig:fraddgp2} but for the nDGP3 simulation, in which the magnitude of the fifth force is much smaller.}
\label{fig:fraddgp3}
\end{figure*}

Figure~\ref{fig:fraddgp3} shows the force profiles from the nDGP3 simulation at $z=1$ and $z=0$. They show the same behavior as in nDGP2, but the fifth force is further suppressed in this model due to the larger value of $r_c$. The ratios in the bottom panel are again constant with radius, such that the voids are completely unscreened, and agree well with the analytic prediction; the values are close to the linear theory values of 0.0314 at $z=1$ and 0.0382 at $z=0$. Though this $r_c$ parameter value produces small differences with respect to \lcdm, it is clear the Vainshtein mechanism is not working at all in cosmic voids, so this could be one of the best places to look to test a large class of modified gravity models.

\subsection{Velocity profiles}

Next we investigate how the velocity profiles trace the fifth forces for voids found using both dark matter particles and halos as tracers of the density field. We measure stacked radial velocity, tangential velocity, and tangential velocity dispersion profiles of cosmic voids and the difference between these profiles in nDGP and \lcdm\ simulations. 

\subsubsection{Radial velocity}

\begin{figure*}
\parbox{0.45\textwidth}{\includegraphics[width=\hsize]{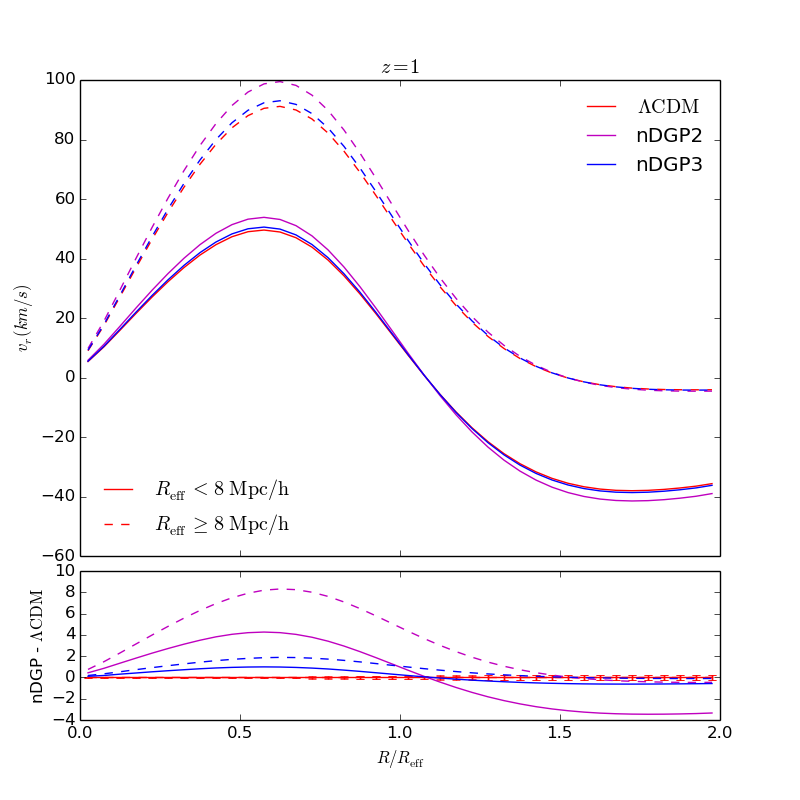}}%
\qquad
\begin{minipage}{0.45\textwidth}%
\includegraphics[width=\hsize]{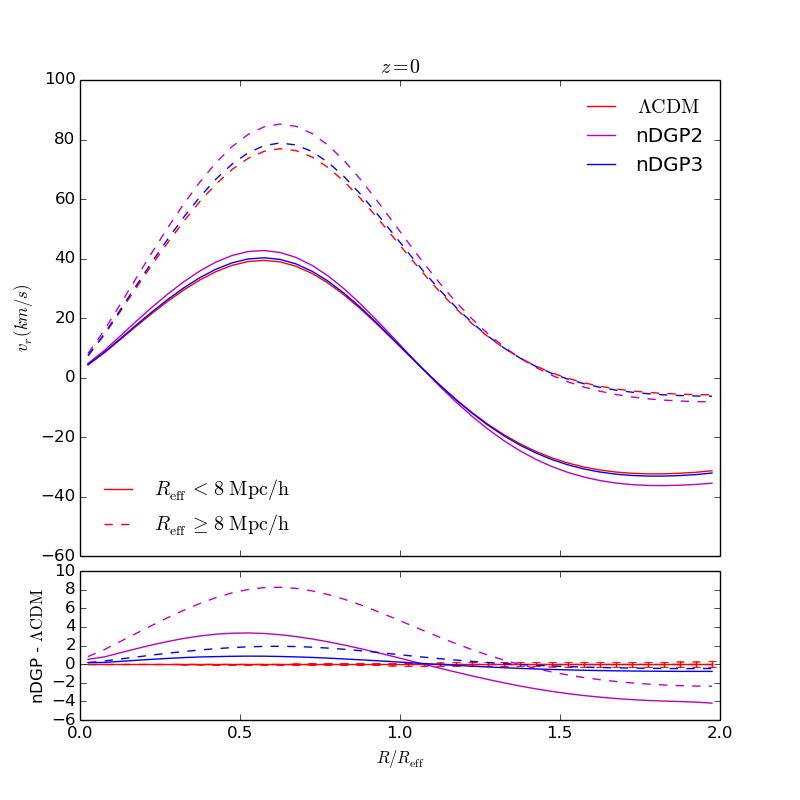}
\end{minipage}%
\caption{Radial velocity profiles for voids found using dark matter particles at $z=1$ (left panel) and $z=0$ (right panel). The difference between the nDGP and \lcdm\ profiles are given in the bottom panels. Error bars (shown for \lcdm\ only) represent the standard deviation of all profiles in the stack.}
\label{fig:vrads}
\end{figure*}

\begin{figure}
\includegraphics[width=\hsize]{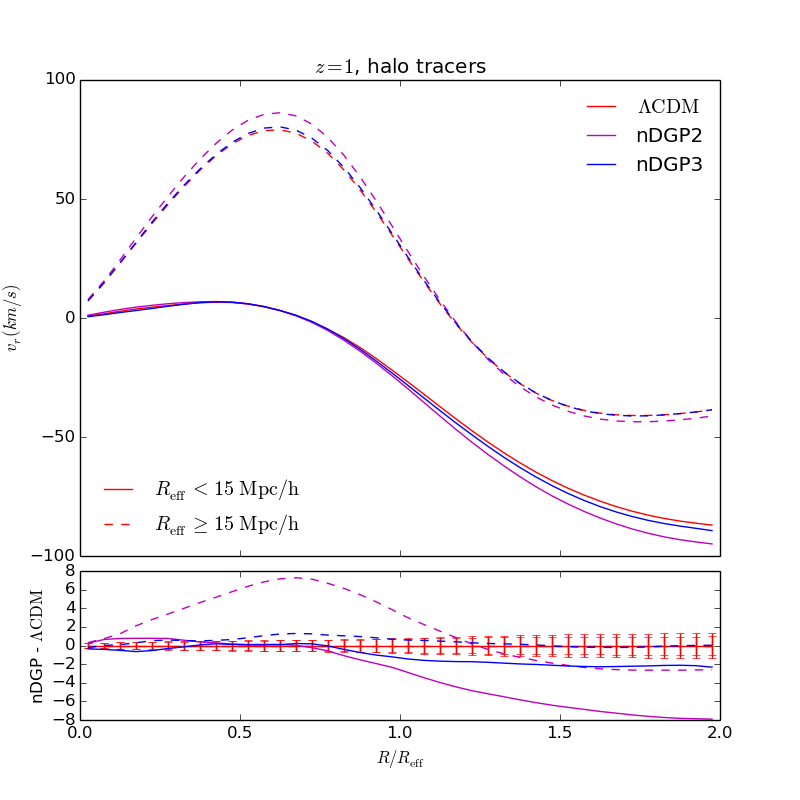}
\caption{Radial velocity profiles: the same as Figure~\ref{fig:vrads} but for voids found in the halo distribution at $z=1$.}
\label{fig:hvrad08}
\end{figure}

Figure~\ref{fig:vrads} shows radial velocity profiles at $z=1$ and $z=0$ for voids found using dark matter particles. As with the radial forces, a positive radial velocity points away from the void centers. The stacked profiles for the smaller sample of voids reach zero at $R=$\reff\ and are negative beyond the void effective radius, as matter flows both out from the void and in toward the local density peak at the void boundary. Radial velocity profiles for the larger voids, however, level off to around zero and don't experience significant infall toward their boundary, and they have a larger magnitude of outflow. The radial velocities of Vainshtein voids are enhanced with respect to \lcdm\ voids, with both larger outflow from void centers and larger inflow. 

We can understand these behaviours qualitatively using the linear theory~\citep{Hamaus2014}. In the linear theory, the radial velocity is given by 
\begin{equation}
v_r(r)= - \frac{1}{3} a f H r \Delta(r) \propto a f F_N(r)
\end{equation}
where $\Delta(r)$ is the integrated density contrast and $f$ is the linear growth rate. As we can see from Figures~\ref{fig:fraddgp2} and~\ref{fig:fraddgp3}, the radial velocity profiles trace the Newtonian force profile. The linear growth rate is enhanced in nDGP, thus it has larger radial velocities. However, this difference is very small, on the order of a few km/s.

The stacked radial velocity profiles of $z=1$ voids found using halo tracers are shown in Figure~\ref{fig:hvrad08}. In contrast to the voids found using dark matter particles, the small voids have very little outfall and are dominated by infall, even within the effective radius. This is in line with the higher central densities of these voids seen in Figure~\ref{fig:haloden08} -- since halos are local peaks in the density field, they are not good tracers of small underdense regions. Though the difference between nDGP and \lcdm\ radial velocity profiles is statistically significant for the nDGP2 model, as seen in the bottom panel of Figure~\ref{fig:hvrad08}, the differences are still very small, on the order of a few km/s, making radial velocities a potentially challenging probe of Vainshtein screening.

\subsubsection{Tangential velocity}

\begin{figure*}
\parbox{0.45\textwidth}{\includegraphics[width=\hsize]{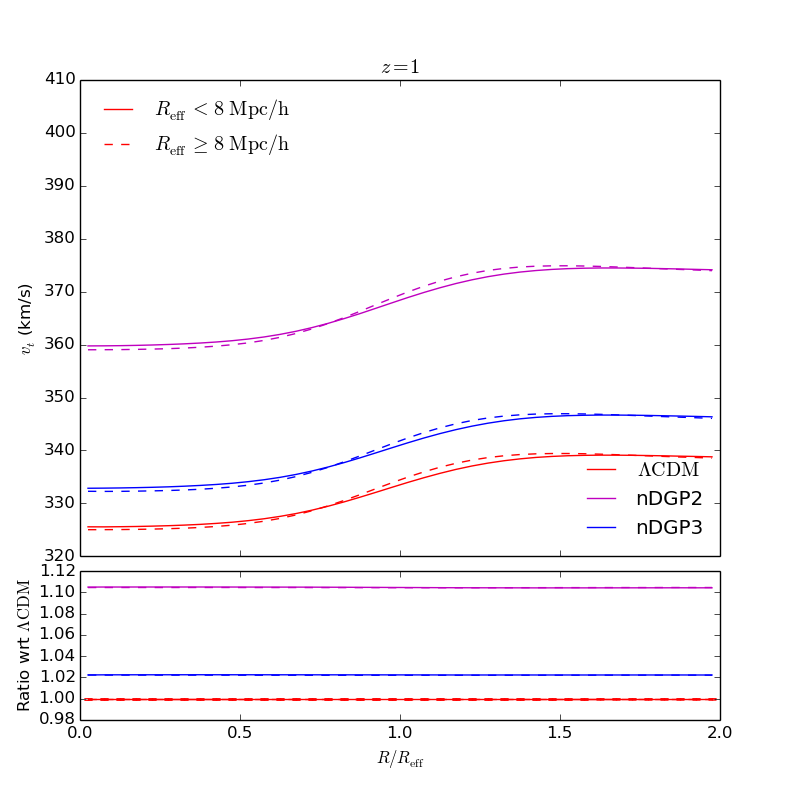}}%
\qquad
\begin{minipage}{0.45\textwidth}%
\includegraphics[width=\hsize]{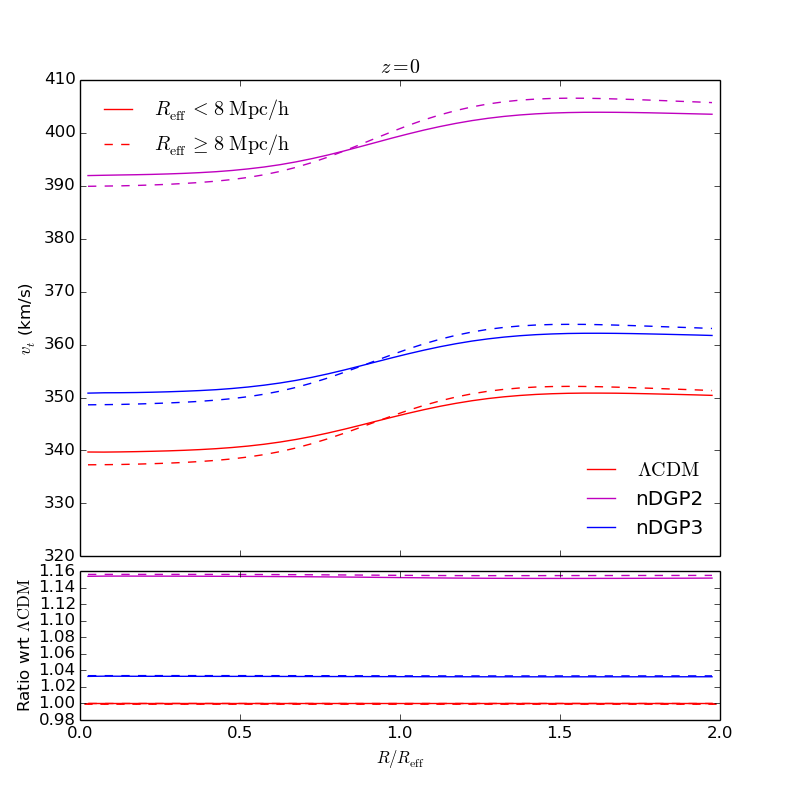}
\end{minipage}%
\caption{Tangential velocity profiles for voids found using dark matter particles at $z=1$ (left panel) and $z=0$ (right panel). The ratio between the nDGP and \lcdm\ profiles are given in the bottom panels. Error bars (shown for \lcdm\ only) represent the standard deviation of all profiles in the stack.}
\label{fig:vtans}
\end{figure*}

\begin{figure}
\includegraphics[width=\hsize]{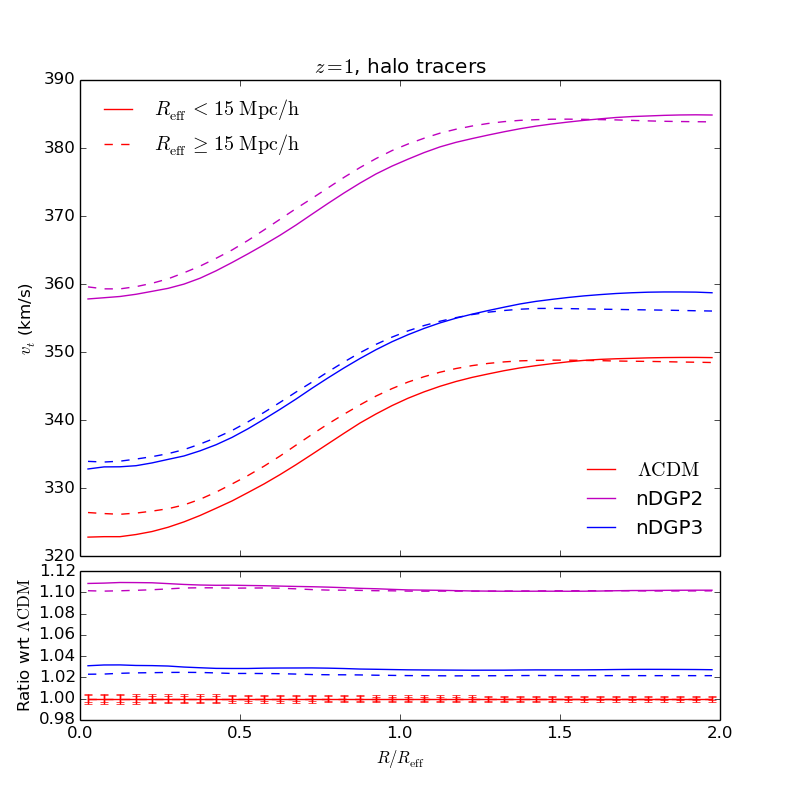}
\caption{Tangential velocity profiles: the same as Figure~\ref{fig:vtans} but for voids found in the halo distribution at $z=1$.}
\label{fig:hvtan08}
\end{figure}

Stacked profiles of the tangential velocity for $z=1$ and $z=0$ are shown in Figure~\ref{fig:vtans}. In contrast to the radial velocity profiles, there is little difference between profiles for the large and small voids and a clear offset with respect to \lcdm\ for the two nDGP models. The weaker dependence of the tangential velocity profiles on void size is a reflection of the fact that the dynamics of large vs. small voids is captured by their outflow from void centres and infall toward void boundaries, not on tangential velocities. The fact that the difference between nDGP and \lcdm\ appears more obvious in tangential versus radial velocity profiles is primarily because the tangential velocities have a much larger magnitude than the velocities radial to the void centres, so the difference between nDGP and \lcdm, which depends on the ratio of the fifth force to Newtonian force, is more apparent. These ratios are given in the bottom panels of Figure~\ref{fig:vtans} and show a constant offset, independent of radius, very near the linear theory values of the ratio of the fifth force to Newtonian force given in Figures~\ref{fig:fraddgp2} and~\ref{fig:fraddgp3}.

The stacked tangential velocity profiles of $z=1$ voids found using halo tracers are shown in Figure~\ref{fig:hvtan08}. Similarly to the voids found using dark matter particles as tracers of the density field, the tangential velocities of halos around voids, using halos as tracers to define the voids, show a clear difference between those in nDGP models and \lcdm. The ratios of the nDGP profiles with respect to \lcdm\ are shown in the bottom panels and again are constant with radius, with values similar to the screening profiles. These profiles show that voids are indeed unscreened in the Vainshtein mechanism and that the velocities around these voids, independent of radius from void centre, are excellent probes of gravity models utilizing Vainshtein screening.

\subsubsection{Velocity dispersion}

\begin{figure*}
\parbox{0.45\textwidth}{\includegraphics[width=\hsize]{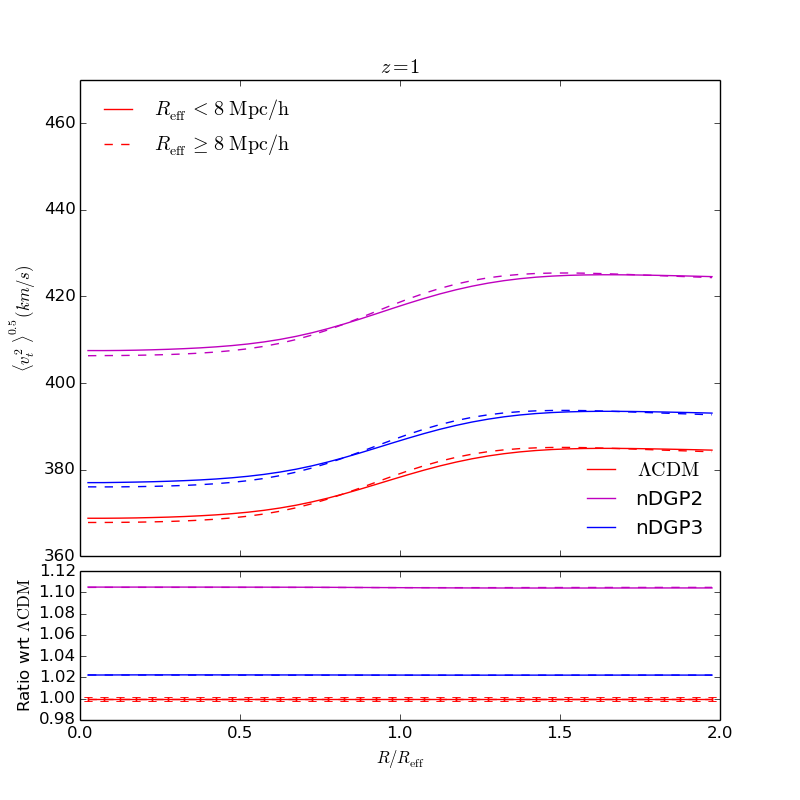}}%
\qquad
\begin{minipage}{0.45\textwidth}%
\includegraphics[width=\hsize]{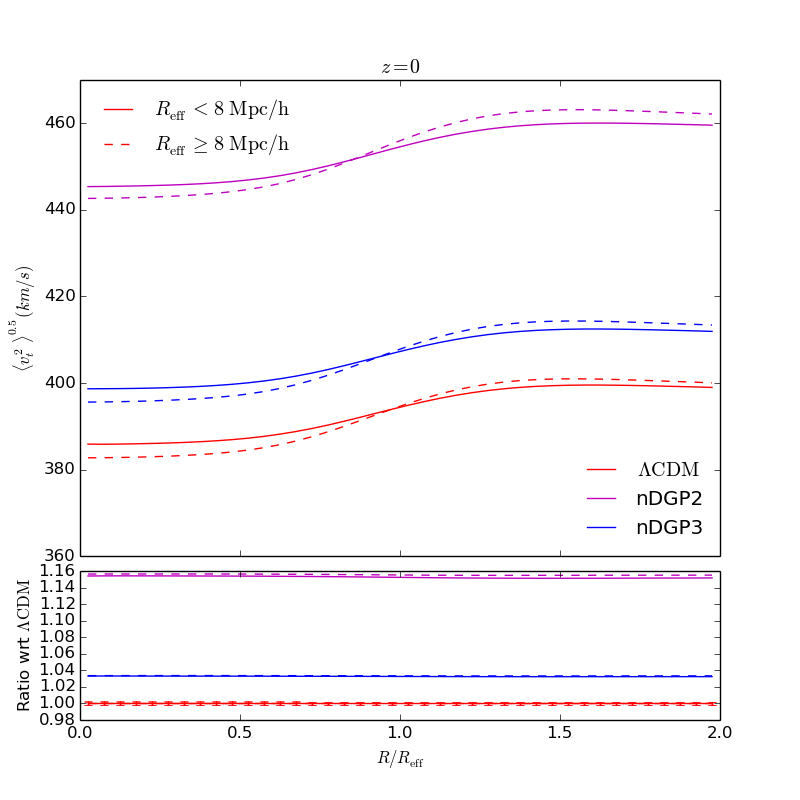}
\end{minipage}%
\caption{Tangential velocity dispersion profiles for voids found using dark matter particles at $z=1$ (left panel) and $z=0$ (right panel). The ratio between the nDGP and \lcdm\ profiles are given in the bottom panels. Error bars (shown for \lcdm\ only) represent the standard deviation of all profiles in the stack.}
\label{fig:v2tans}
\end{figure*}

\begin{figure}
\includegraphics[width=\hsize]{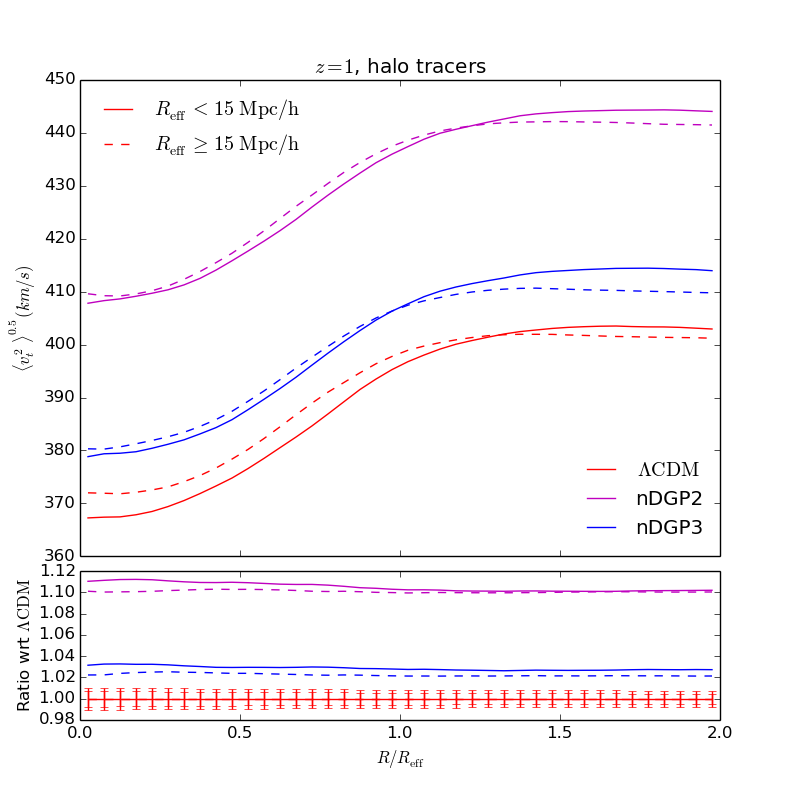}
\caption{Tangential velocity dispersion profiles: the same as Figure~\ref{fig:v2tans} but for voids found using halo tracers at $z=1$.}
\label{fig:hv2tan08}
\end{figure}

Velocity dispersion profiles around dark matter halos and moments of the pairwise velocity dispersion have been suggested as probes of modified gravity in $f(R)$ and Galileon models~\citep{Lam2012,Hellwing2014}. The dispersion profile around halos and clusters must be probed far into the outer regions where the screening is no longer damping the signal. We have seen that voids are not screened at all by the Vainshtein mechanism, and indeed this is reflected in the profiles of the tangential velocity dispersion, shown in Figure~\ref{fig:v2tans} for voids found in the dark matter particle distribution at $z=1$ and $z=0$. There is little difference between profiles for the large and small samples, while the nDGP profiles show a constant offset with respect to \lcdm; ratios of nDGP with respect to \lcdm\ are shown in the bottom panels. As with the tangential velocity profiles, the values of these ratios are remarkably close to the ratios of the fifth force to Newtonian force found in the screening profiles (Figures~\ref{fig:fraddgp2} and~\ref{fig:fraddgp3}) for a given simulation and redshift.

It is interesting to note that for $f(R)$ models of gravity that experience chameleon screening, the offset of the velocity dispersion profiles of $f(R)$ with respect to \lcdm\ voids was not found to be constant but was smaller within the void and increased at about the void radius~\citep{Cai2015}. However, \citet{Cai2015} use a different void finder for their study, so it is not clear whether this difference is caused by the difference in the screening mechanisms or different definition of voids.

Very similar results are found when halos are used to define the voids and measure the velocity dispersion profiles, shown in Figure~\ref{fig:hv2tan08} for $z=1$. The dispersions themselves have lower values than in voids found using dark matter particles (Figure~\ref{fig:v2tans}), but the ratios of nDGP to \lcdm\ are similar and again closely match the screening profiles, though the error bars on the profiles are larger because there are fewer voids. Though there are on the order of 10,000 voids that have been stacked to compute these profiles, a void catalog of this size is within reach of current surveys~\citep{Nadathur2016DR11}. However, in order to distinguish between \lcdm\ and a model with Vainshtein screening, the velocities would have to be measured very well, as the differences can be $\sim$ 10 km/s.


\section{Conclusion}
\label{sec:conc}

The Vainshtein mechanism, which appears in many models of modified gravity, is very effective at screening dark matter halos but not the other elements of the cosmic web, including voids~\citep{Falck2014}. We have investigated the effect of the Vainshtein mechanism on the properties of cosmic voids in cosmological $N$-body simulations by measuring their density, fifth force, velocity, and velocity dispersion profiles. The voids are identified with a watershed technique, using both dark matter particles and dark matter halos as tracers of the density field, at $z=1$ and $z=0$.

The density profiles of stacked voids show that Vainshtein voids are emptier in the centres compared to voids in \lcdm, similarly to what was found for $f(R)$ and Galilieon voids~\citep{Cai2015,Barreira2015}. At $z=1$, Vainshtein voids have $\sim 5$\%-1\% deeper void centers, depending on the strength of the modification to gravity, and the difference with respect to \lcdm\ levels off at the effective void radius. These differences in void centers increase to $\sim 10$\%-2\% at $z=0$. In contrast to the density profiles, the fifth force profiles show a constant force ratio at all radii, with a value that is at the level of the linear theory value. Thus Vainshtein voids are unscreened, independent of void radius, out to as far as twice the void effective radius. This holds for voids found both using dark matter particles and halos as tracers of the density field. Both the density and fifth force profiles showed similar results for the two sets of large and small voids: even though many small voids are likely in dense environments, i.e. voids-in-clouds~\citep{Sheth2004}, these overdense `clouds' are not collapsing halos and thus Vainshtein screening is not triggered~\citep{Falck2014}.

The radial velocities of Vainshtein voids are enhanced with respect to \lcdm\ voids, with both larger outflow from void centres and larger inflow, especially for small voids. However, the magnitude of this difference is very small, on the order of a few km/s for both $z=1$ and $z=0$. On the other hand, tangential velocity and velocity dispersion profiles show a clear offset between Vainshtein and \lcdm\ voids. The ratios between Vainshtein and \lcdm\ tangential velocity and velocity dispersion profiles show a constant offset as a function of void radius, with values that are very close to the linear theory values of the fifth force to Newtonian force ratios. This is true at both $z=1$ and $z=0$, using both dark matter particles and halos as tracers to define the voids, though the statistical error is increased for halo-tracer voids because there are fewer of these voids. This suggests that tangential velocities of voids are excellent tracers of the enhanced fifth force in models of gravity that exhibit the Vainshtein screening mechanism. 

Observing this signature, which is tangential to the void centers, would require measuring the velocity dispersion of galaxies on the edges of voids with respect to our line-of-sight, which is where most void tracers should be anyway; however, a full accounting of this effect would need to take into account the effect of redshift space distortions~\citep[see, e.g.][]{Hamaus2016,Cai2016,Hawken2016}. Further, void catalogs derived from surveys use galaxies as tracers of the density field, which are more highly biased and sparser tracers than halos, leading to a smaller sample of identified voids with different size distributions than those found using dark matter ~\citep{Nadathur2015Voids2}. However, since we have shown Vainshtein voids to be completely unscreened for all void sizes, this will primarily affect that statistical significance of the signal. Thus, the stacked tangential velocity profiles of a large sample of voids is a promising signature of Vainshtein models of gravity.

\section*{Acknowledgements}

BF acknowledges support from the Research Council of Norway (Programme for Space Research). KK is supported by the European Research Council through 646702 (CosTesGrav) and the UK STFC grant ST/N000668/1. GBZ is supported by the 1000 Young Talents program in China and by the Strategic Priority Research Program ``The Emergence of Cosmological Structures'' of the Chinese Academy of Sciences, Grant No. XDB09000000. 
MC is supported by ERC Advanced Investigator grant COSMIWAY (grant number GA 267291) and the UK STFC grant ST/L00075X/1. 
Numerical computations were done on the Sciama High Performance Compute (HPC) cluster which is supported by the ICG, SEPNet, and the University of Portsmouth.


\bibliographystyle{mnras}
\bibliography{vainshteinvoids}


\bsp	
\label{lastpage}
\end{document}